# The Retrieval Effectiveness of Web Search Engines: Considering Results Descriptions[1]

*Dirk Lewandowski*, Hamburg University of Applied Sciences, Faculty Design, Media and Information Department Information, Berliner Tor 5, D – 20249 Hamburg, Germany.
E-Mail: dirk.lewandowski@haw-hamburg.de, Tel: ++49 40 42875-3621, Fax: ++49 40 42875-3609

**Abstract**

Purpose: To compare five major Web search engines (Google, Yahoo, MSN, Ask.com, and Seekport) for their retrieval effectiveness, taking into account not only the results but also the results descriptions.
Design/Methodology/Approach: The study uses real-life queries. Results are made anonymous and are randomised. Results are judged by the persons posing the original queries.
Findings: The two major search engines, Google and Yahoo, perform best, and there are no significant differences between them. Google delivers significantly more relevant result descriptions than any other search engine. This could be one reason for users perceiving this engine as superior.
Research Limitations: The study is based on a user model where the user takes into account a certain amount of results rather systematically. This may not be the case in real life.
Practical Implications: Implies that search engines should focus on relevant descriptions. Searchers are advised to use other search engines in addition to Google.
Originality/Value: This is the first major study comparing results and descriptions systematically and proposes new retrieval measures to take into account results descriptions.
Article type: Research paper
Keywords: Word Wide Web / search engines / retrieval effectiveness / results descriptions / retrieval measures

**Introduction**

It is hard to say which of the major international search engines is the best. One reason is that there are a variety of quality factors that can be applied to search engines. These can be grouped into four major areas (Lewandowski & Höchstötter, 2007):

1. Index Quality: This points out the importance of the search engines' databases for retrieving relevant and comprehensive results. Areas of interest include Web coverage (e.g., Gulli & Signorini, 2005), country bias (Vaughan & Thelwall, 2004), and up-to-dateness (Lewandowski, Wahlig, & Meyer-Bautor, 2006).
2. Quality of the results: This is where derivates of classic retrieval tests are applied. However, it should be asked which measures should be applied and if new measures are needed to satisfy the unique character of the search engines and their users (Lewandowski, 2007a).
3. Quality of search features: A sufficient set of search features and a sophisticated query language should be offered and work reliably.
4. Search engine usability: Here it is asked whether it is possible for users to interact with search engines in an efficient and effective way.

The study presented in this article deals solely with the quality of the results. However, it should be kept in mind that while the results quality is important, it is only one area where the quality of search engines can be measured. In further research, we hope to present a balanced quality model that takes into account all the factors mentioned.

---

[1] This is a preprint of an article accepted for publication in the Journal of Documentation (2008).

The area of search engine quality research derives its importance not only from a general interest of information science in the performance of search systems but also from a wider discussion on search engines in general society. Market figures show that there is a dominance of Google, at least on the European market. While in the US, Google serves (according to different sources) around 50 to 60 percent of all Web search queries (Burns, 2007; Sullivan, 2006); the situation in Europe is quite different. In all major European countries, Google has a market share of more than 80 percent (e.g., for Germany, see Webhits Web-Barometer, 2006). This led to a discussion whether there should be a European search engine countering the dominance of Google, or, more generally, of the US-based search engines. However, it is interesting to see that there is a wide knowledge gap on the performance of these engines. Does Google dominate the European search market, because it provides the users with the best results? Or are users just used to Google, while other search engines provide at least comparable results? The conclusions from these two scenarios would be quite different. If Google were the only engine to produce high-quality results, then the user would be right to use this engine exclusively. It can be concluded that building another engine with better results could lead to a certain amount of market share.

If the other scenario were true, the users' preference for Google would be just a case of market shares. If users only knew about other, comparable search engines, they could be willing to switch their favourite search engine.

There are some studies on the retrieval effectiveness of search engines, which will be reviewed in the literature section. We found two major reasons to conduct our own test: Firstly, the search engine landscape has changed significantly since the last investigations. Secondly, we found that the importance of the results descriptions was not adequately taken into account in most of the studies.

When comparing the search engines from a purely user-centred perspective, one finds that users are satisfied with their favourite search engine (Machill, Neuberger, Schweiger, & Wirth, 2003 pp. 175-188) whether it is Google or not. However, in the case of Google, preference for the favourite engine is stronger than for any other search engine. However, looking at the search engines users know about, we see that users generally know only about one or a few search engines and do not compare the results given by these. Once the favourite search engine is chosen, they usually stick to it.

With our study, we hope to find answers to the following questions:

1. Which search engine performs best in terms of precision? Do search engines now perform better than some years ago, when the last studies on their retrieval effectiveness were conducted?
2. Are the results given by the search engines properly described, i.e. do relevant result descriptions lead to relevant results?
3. How can the results descriptions be taken into account in relevance measures?
4. Do overall relevance judgements change when not only the results, but also the results descriptions are considered?

**Literature Overview**

This section focuses on literature on the retrieval effectiveness of Web search engines. General literature on retrieval effectiveness as well as a general discussion on the concept of relevance is omitted. For our overview, we chose retrieval effectiveness studies that are interesting due to their methodology and their results, respectively. Studies range from 1996 to 2007, but we do not intend to give a complete overview of such studies. Criteria discussed in this section are based on recommendations for the design of general retrieval tests (Tague-Sucliffe, 1992) as well as for Web-specific evaluations (Gordon & Pathak, 1999; Hawking, Craswell, Bailey, & Griffiths, 2001).

**Table I** Overview of search engine retrieval effectiveness tests

| Authors | Year | Query language | Number of queries | Number of results | Number of engines tested | Engines tested | Query topics | Jurors | Information need stated? | Relevance judged by actual user? | Relevance scale | Source of results made anonymous? | Results lists randomised? |
|---|---|---|---|---|---|---|---|---|---|---|---|---|---|
| Chu & Rosenthal | 1996 | English | 10 | 10 | 3 | AltaVista, Excite, Lycos | Queries drawn from real reference questions | Persons conducting the test | no | no | scale (0; 0.5; 1) | no | no |
| Ding & Marchionini | 1996 | English | 5 | 20 | 3 | Infoseek, Lycos, OpenText | 3 randomly selected from a question set for Dialog online searching exercises in an information science class, 2 personal interest | Person conducting the test | no | no | scale (6 point) | no | no |
| Leighton & Srivastava | 1999 | English | 15 | 20 | 5 | AltaVista, Excite, HotBot, Infoseek, Lycos | reference queries at a University library | Person conducting the test | yes | no | categories (4) + duplicate link + inactive link | yes | |
| Gordon & Pathak | 1999 | English | 33 | 20 | 8 | AltaVista, Excite, Infoseek, OpenText, HotBot, Lycos, Magellan, Yahoo Directory | business-related | Faculty from a business school | yes | yes | scale (4 point) | no | yes |
| Wolff | 2000 | German | 41 | 30 | 4 | AltaVista, C4, MetaCrawler, NorthernLight | 20 general interest, 21 scientific | | no | yes | scale (3: rel, non rel, maybe rel) | yes | ? |
| Dresel et al. | 2001 | German | 25 | 25 | 6 | Abacho, Acoon, Fireball, Lycos, Web.de directory, Yahoo directory | five groups: products, guides, scientific, news, multimedia | Person conducting the test | no | no | binary | no | no |
| Griesbaum et al. | 2002 | German | 56 | 20 | 4 | Google, Lycos, AltaVista, Fireball | scientific; society | students and faculty | no | no | binary/scale (3 point) | yes | no |
| Griesbaum | 2004 | German | 50 | 20 | 3 | Google, Lycos, AltaVista | general interest | students | no | no | binary/scale (3 point) | yes | no |
| Véronis | 2006 | French | 70 | 10 | 6 | Google, Yahoo, MSN, Exalead, Dir, Voila | 14 topic areas; general interest | students | no | yes | scale (0-5) | yes | yes |
| Lewandowski (current study) | 2007 | German | 40 | 20 | 5 | Google, Yahoo, MSN, Ask, Seekport | general interest | students | yes | yes | binary, scale (1-5), scale (0-100) | yes | yes |

Table I shows the overview while we will discuss the major points in the text (for an overview of older tests, see Gordon & Pathak, 1999, p. 148).

While most tests use precision as a retrieval measure, there are some studies that use other metrics. First, we have to group the tests into categories according to their research intent. There are studies that test search engines for availability of Web documents (Stock & Stock, 2000) or their ability to retrieve homepages (Hawking & Craswell, 2005). However, we want to focus on studies that deal with result sets and not with queries that should be answered with just one result. We will discuss the different query types extensively in the methods section.

We only review studies that either use English or German language queries. One exception is the Véronis' study (2006), which uses French queries. We included it in our overview due of its methodology and its topicality.

The number of queries used in the studies varies greatly. Especially the oldest studies (Chu & Rosenthal, 1996; Ding & Marchionini, 1996; Leighton & Srivastava, 1999) use only a few queries (5 to 15) and are, therefore, of limited use (see Buckley & Voorhees, 2000). Newer studies use at least 25 queries, some 50 or more.

In older studies, queries are usually taken from reference questions or commercial online systems, while newer studies focus more on the general users' interest or mix both types of questions. There are studies that deal with a special set of query topics (e.g., business, see Gordon & Pathak, 1999), but we see a trend in focusing on the general user in search engine testing.

Most studies use students as jurors. This comes as no surprise as most researchers teach courses where they have access to students that can serve as jurors. In some cases, the researchers themselves are the jurors (Chu & Rosenthal, 1996; Ding & Marchionini, 1996; Dresel et al., 2001; Griesbaum, 2004; Griesbaum, Rittberger, & Bekavac, 2002; Leighton & Srivastava, 1999).

Regarding the number of results taken into account, most investigations only consider the first 10 or 20 results. This has to do with the amount of work for the evaluators but also with the general behaviour of search engine users. These users only seldom view more than the first one or two results pages (with usually 10 results on each page). Therefore, a cut-off value of 10 or 20 appears reasonable. The selection of search engines to investigate follows clear decisions. Usually, the biggest and most popular search engines are tested, sometimes in comparison to newer or language-specific search engines (e.g., Véronis, 2006).

An important question is how the results should be judged. Most studies use relevance scales (with three to six points). Griesbaum's studies (2004, 2002) use binary relevance judgements with one exception: results can also be judged as "pointing to a relevant document" (i.e., the page itself is not relevant but has a hyperlink to a relevant page). This is done to take into account the special nature of the Web. However, it seems problematic to judge these pages as (somehow) relevant, as pages could have many links, and a user then (in bad circumstances) has to follow a number of links to access the relevant document.

Two important points in the evaluation of search engine results are whether the source of results is made anonymous (i.e., the jurors do not know which search engine delivered a certain result) and whether the results lists are randomised. The randomisation is important to avoid learning effects.

While most of the newer studies do make the results anonymous, we found only two studies that randomise the results lists (Gordon & Pathak, 1999; Véronis, 2006).

While the older studies are mainly interesting for their methodological approaches, we also want to report the findings of some of the newer studies. Griesbaum (2004) finds that Google outperforms Lycos and AltaVista, but the precision values at 20 results for all search engines do not differ much (Google: 0.59, Lycos: 0.53, AltaVista: 0.51). The results show that the overall precision of all search engines is rather low.

Véronis (2006) uses a five-point relevance scale to judge results. This study also shows that neither of the engines tested receives a good overall relevance score. The author concludes that "the overall grades are extremely low, with no search engine achieving the 'pass' grade of 2.5" (Véronis, 2006, p. 7). The best search engines are Yahoo and Google (both 2.3), followed by MSN (2.0). The other (French) search engines perform worse with 1.8 for Exalead, 1.4 for Dir, and 1.2 for Voila.

The results from these studies show that search engines have problems in producing satisfying results lists.

Some work has been done on evaluating search engines' retrieval effectiveness for queries in other languages than English. A good overview of these studies can be found in Lazarinis (2007). As we use German language queries for our investigation, it would be interesting to know if there are differences in the search engines' performance for German and English queries in general. However, it is hard to decide whether the performance of the international search engines differ for English and German language queries. While Lazarinis (2007) found that the engines do have problems with diacritics and other language-specific attributes, this appears not to be true for the German language (Guggenheim & Bar-Ilan, 2005). There is no research yet that directly compares the performance of the engines on English versus German queries. It would be interesting to see whether the average precision values are generally lower for German queries.

*The Importance of Results Descriptions*

The importance of the results descriptions derives from the two steps in the searching process. First, a user chooses documents from the search engine results pages, then evaluates the documents themselves. These predictive judgements lead to viewing certain documents. These documents—and only these—are judged for relevancy (evaluative judgement) (Crystal & Greenberg, 2006). However, in conventional retrieval effectiveness studies, only the results themselves are evaluated. Predictive judgements are omitted, although these form an integral part of the user's searching process.

What information should a good result description provide? Crystal and Greenberg (2006, p. 1370) state that "Good surrogates should provide metadata that enable users to predict the relevance of the document quickly and accurately." According to these authors, the two main questions in generating results descriptions are: How much metadata is useful? And which metadata elements are particularly valuable?

The amount of metadata used is relatively consistent with different search engines. Result descriptions are relatively short. Google, Yahoo, and Ask.com only use one or two lines of text with up to about 20 words, while MSN and Seekport use longer results descriptions.

In the composition of results descriptions, search engines follow different approaches, as well. While the methods actually used vary from one engine to another, there are just a few possibilities used by search engines in general. First, results descriptions can be composed as keywords in context (KWIC). This is the method used for most of the results descriptions, because data from the results themselves is available for all hits. The disadvantage of this method is that KWIC often shows incomplete sentences, and it is, therefore, difficult for the user to judge the results.

The second method uses external data from directories. Some search engines use Website descriptions from the Open Directory Project (ODP); Yahoo uses descriptions from its own Web directory for some results. The advantage of using directory data is that in an ideal case, the search engine would give a concise description and the user could judge the results accordingly. However, the disadvantage of this method lays in the inadequate availability of directory data, which is only available for complete Websites, not individual pages. In addition, directories are far from being complete, and so this kind of data can only be used for popular Websites that are included in the directories.

The third data source for results descriptions is metadata from the pages themselves. Where such data is available, it often gives a useful description of the page. However, here as well, data is only available for some pages.

An approach that was used by some search engines in the nineties was building results descriptions from the first sentences of the documents. We mention this approach just for completeness here, because it is not used by any of the major search engines anymore.

**Figure 1** Results description for the Windows update site from different search engines
(from top: Google, Yahoo, Live.com, Ask.com)

Figure 1 shows the results descriptions for the Windows Update site as an example for the use of different approaches by the different search engines. Google uses the contents of the META description tag; Yahoo uses the description from its own directory of Web sites, while MSN and Ask.com use (a part of) the description from the Open Directory.

While the presentation of search results in commercial databases is discussed in some articles (e.g., Jacsó, 2005), we found only two studies that used results descriptions for measuring search engines' performance. Wolff (2000) solely uses results descriptions to compare four (meta) search engines. In Griesbaum's 2004 study, the effectiveness of results descriptions is compared to the actual results. He finds that with about one third of all results, the judgement of the description differs from the judgement of the result itself. In some cases, results appear better than their descriptions, in some cases vice versa. Griesbaum notes that his results are of a tentative nature. The main concern against the results is that different jurors judged descriptions and results.

**Methods**

*Choice of Search Engines*
For our investigation, five search engines were chosen. These are Google, Yahoo, MSN [1], Ask.com, and Seekport. Our first criterion was that each engine should provide its own index of Web pages. Many popular search portals (such as AOL) make use of results from the large search providers. For this reason, these portals were excluded from our study. The second criterion was the popularity of the search services. According to Search Engine Watch, a search industry news site, the major international search engines are Google, Yahoo, Microsoft's Live.com, and Ask.com (Sullivan, 2007). Therefore, these were chosen for further investigation. Furthermore, we added the German search engine, Seekport, to our study. Seekport is the only noticeable German search engine that provides its own index.

*Participants*

Participants of this study were recruited from a student group at the Heinrich-Heine-University Düsseldorf. In total, 40 persons participated in the study. First, they had to write down a query, which later was used for testing the search engines. Every participant had to judge the relevance of the results of her own query, but results were made anonymous so that participants were not able to see from which search engine a particular result was and at which rank it originally was shown.

*Queries*

The choice of queries was key to our evaluation. While search engines have to serve different kinds of queries, our study only deals with informational queries.

According to Broder (2002), with *informational queries*, users want to find information on a certain topic. Such queries usually lead to a set of results rather than just one suitable document. Informational queries are similar to queries sent to traditional text-based information retrieval systems. According to Broder, such queries always target static Web pages. However, the term *static* here should not refer to the technical delivery of the pages (e.g., dynamically generated pages by server-side scripts such as php or asp) but rather to the fact that once the page is delivered, no further interaction is needed to get the desired information.

In opposition to that, navigational queries are used to find a certain Web page the user already knows about or at least assumes that such a Webpage exists. Typical queries in this category are searches for a homepage of a person or organization. Navigational queries are usually answered by just one result; the informational need is satisfied as soon as this one right result is found.

The results of transactional queries are Web sites where a further interaction is necessary. A transaction can be the download of a program or file, the purchase of a product, or a further search in a database.

Based on a log file analysis and a user survey (both from the AltaVista search engine), Broder finds that each query type stands for a significant amount of all searches. Navigational queries account for 20-24.5 percent of all queries, informational queries for 39-48 percent and transactional queries for 22-36 percent. Later studies (e.g., Rose & Levinson, 2004) confirm that all three query types account for a significant ratio of total queries.

These figures show that our study cannot make general statements on the effectiveness of search engines but only on their performance with informational queries. While other query types should be considered in separate research, we think that query types should not be mixed in one investigation, because relevance judgments for each type should be different.

For example, Griesbaum uses all three query types in his 2004 study. We find informational queries as well as navigational queries ("hotmail", "yahoo") and transactional queries ("disneyworld tickets"). This makes it difficult to judge the results. For the navigational queries, we cannot assume that there are 20 relevant results (the cut-off value used in the study). Therefore, results are biased in some way.

For our investigation, we use 40 queries. This number results from the number of participants we were able to obtain. As the judgment of the results was a very work-intensive task, we were not able to motivate more participants. Buckley & Voorhees (2000) suggest using 50 queries but define the lower limit to 25 queries. Our query number is well within this scale.

Participants were asked to write down the last query they posed to a public search engine. We asked for queries used in their private lives, not used for research for their studies. All queries were German language. In addition, we asked the participants to provide short descriptions of their information need underlying the query.

We hoped to get queries that represent general user behaviour at least to some degree. Table II shows the distribution of query length for the queries used in this study. The average length of the queries used is 2.3 terms, while in general log file based investigations, the average query length lies between 1.7 and 1.8 terms (Schmidt-Maenz & Koch, 2006).

**Table II** Query length

| Complexity | Number of queries | Percent |
|---|---|---|
| 1 word | 13 | 32.5 |
| 2 words | 12 | 30.0 |
| 3 words | 7 | 17.5 |
| 4 words | 3 | 7.5 |
| 5 words | 5 | 12.5 |

We also analysed our queries for search topics. Table III shows the distribution of topics in our data set versus the distribution of topics from a general investigation of the search topics of German Web search engine users. The classification scheme used is from (Spink, Wolfram, Jansen, & Saracevic, 2001). When comparing this distribution with the distribution of general German language Web search queries, we find that our queries do not exactly fit with the general queries. Education and society are overrepresented, while the most popular topic with the general users, commerce, is underrepresented in our data set. However, with a relatively small number of queries, one could not expect a better compliance. The data show that (with the exception of sex and government) our queries cover all major topics.

As could be expected in our research environment, our queries are more complex than the average queries of German Web search engine users. In addition, we have a different topic distribution. Nevertheless, we think that our query set is near enough to the general user behaviour to produce valid results.

**Table III** Query topics of the current study compared to general topic distribution of German queries

| Topic* | Number of queries | Percent | Percentage of general German language Web queries** |
|---|---|---|---|
| Commerce, travel, employment, or economy | 11 | 27.5 | 36.5 |
| People, places or things | 3 | 7.5 | 16.1 |
| Entertainment or recreation | 5 | 12.5 | 9.7 |
| Computers or Internet | 5 | 12.5 | 9.3 |
| Sex or pornography | 0 | 0 | 5.7 |
| Health or sciences | 3 | 7.5 | 9.2 |
| Education or humanities | 5 | 12.5 | 2.6 |
| Government | 0 | 0 | 4.3 |
| Performing or fine arts | 2 | 5 | 1.5 |
| Society, culture, ethnicity, or religion | 6 | 15 | 5.0 |

\* Classification from (Spink et al., 2001)
\*\* from (Lewandowski, 2006). Queries not classifiable were omitted.

*Number of Results*
For each query, we analysed the first 20 results from each search engine. A cut-off value had to be used because of the very large results sets produced by the search engines. From user studies, we know that users only look at the first few results and ignore the rest of the result list. We used 20 results per search engine to make our results comparable to older studies and to have enough data to compare only results with relevant results descriptions, as well.

*Relevance Judgements*
While there is a long and ongoing discussion on relevance in information science (see overviews in Borlund, 2003; Mizzaro, 1997), we wanted to restrict relevance judgements to a binary decision whether a result (or a result description, respectively) is relevant or not. We collected different relevance judgements from our jurors (binary and two different scales), but in this study, we only want to use binary judgements. In further work, we will compare the values from the different scales.
In the case of the results, users were asked to state whether the document was relevant to the query and the underlying information (as expressed in the questionnaire) or not. We decided against using graded relevance for results that are not relevant themselves, but are pointing to a relevant result as used in Griesbaum (2004) and Griesbaum et al. (2002). We think that a search engine should provide direct links to the relevant results and that users are not satisfied with one further step in the searching process.
Relevance judgements in case of the descriptions were more complicated. A description should be relevant if the user thinks that this surrogate points to a relevant result. Therefore, participants were asked to judge the descriptions in this respect and not in respect to whether the description itself contains information that helps satisfy the information need.

*Data Collection*

The data for this study was collected on the 20th and 21st of January 2007. Participants were given one query (not their own) and asked to submit it to the search engines under investigation. For each result [2], participants were asked to copy the results description and the results URL into a Microsoft Excel sheet. When this process was finished for all queries, the data was made anonymous, so that participants were not able to see which result was from which search engine. Additionally, results were randomised, so that the original order of the results had no influence on the relevance judgements.

After this, the Excel sheet was given to the participant who originally submitted the query for investigation. In the first step, the user was only able to see the results descriptions and was asked to judge these descriptions on the basis of whether it would lead to a relevant result.

In the second step, participants were given the results URL, but were not able to see the results descriptions anymore. They were again asked to judge the relevance of each hit.

**Results**

In this section, we discuss the precision of the results on a micro and macro level. Then we focus on the results descriptions and the differences between the precision of descriptions and results. Then, we develop new measures that combine the relevance of the descriptions with the ones of the results.

*Number of Relevant Results*

Figure 2 shows the relative number of relevant result descriptions and results for each search engine. We used relative values, because not every search engine produced our maximum of 20 results for every query. Therefore, absolute values would not have been meaningful.

As expected, the values for the descriptions are higher than for the results themselves. Regarding only the results, Yahoo comes first with 48.5 percent relevant results, followed by Google with 47.9 percent. However, a chi square test shows that these differences are not significant, while differences between all other engines are highly significant ($p<0.01$). The results show that while Google is usually regarded as the best search engine, its results are not better than Yahoo's.

On the other end of the spectrum, the most surprising result is that the Microsoft's MSN search engine does not perform much better than Seekport, a search engine run by a small German company. It is not the good performance of Seekport that should be emphasised here, but the disappointing performance of MSN.

When looking at the results descriptions, we find that Google performs best with 60.2 percent of descriptions judged as leading to a relevant result. Yahoo comes second with 52.8 percent. For the other three engines, the ratio of relevant result descriptions is well under 50 percent. Differences between all search engines are highly significant according to the chi square test.

The number of relevant results descriptions could at least in part explain the perception of Google as the engine with the best results. As users are not able to systematically compare the results of the different engines, they have to rely on their impression of the results, which largely comes from the results lists. A user sees many more descriptions than results when using search engines and only picks the results that are judged relevant from the results descriptions.

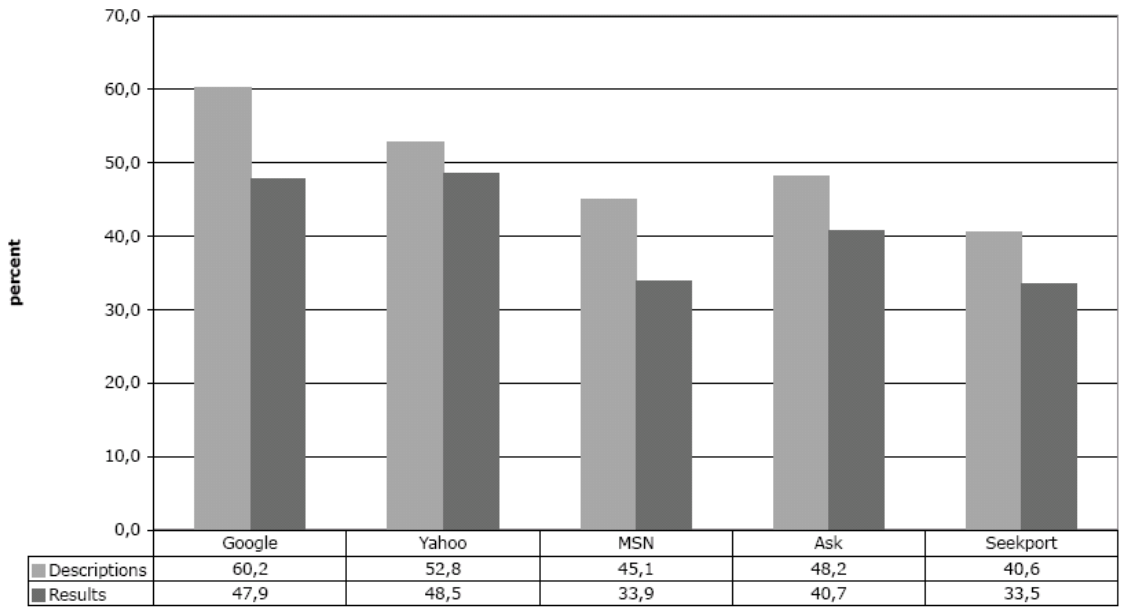

**Figure 2** Comparison of search engines on the ratio of relevant results descriptions vs. relevant results

We further asked whether the search engines are capable of producing at least one relevant result for each query. For a searcher, the greatest disappointment would be to get such a result set. However, the search engines investigated produce good results in this respect. While Google and Yahoo are able to answer all queries with at least one relevant result, we found Ask.com did not answer one query at all, and MSN and Seekport did not answer three queries from our set.

*Micro Precision*
While the number of relevant results produced does not take into account whether a result is presented on the first or last ranking position under investigation, micro precision does exactly that. In Figure 3, the distribution of results is plotted. The graphs show how the relevant results spread among the single positions. It can be read from the graph how good an individual search engine performs when taking into account only the top x results. For example, one can see that the search engines perform differently when considering only the top 3 results or the complete set of 20 results. For the top positions, one can see clearly that Google and Yahoo perform best. The difference to Ask.com is relatively high when comparing to the difference for the whole set of 20 results. This indicates that while Ask.com produces good results for the whole results set, ranking should be optimised to get better results on the first few ranking positions. These are mainly viewed by the users (Granka, Joachims, & Gay, 2004). Therefore, search engines should take care that the first few results are of high relevance.

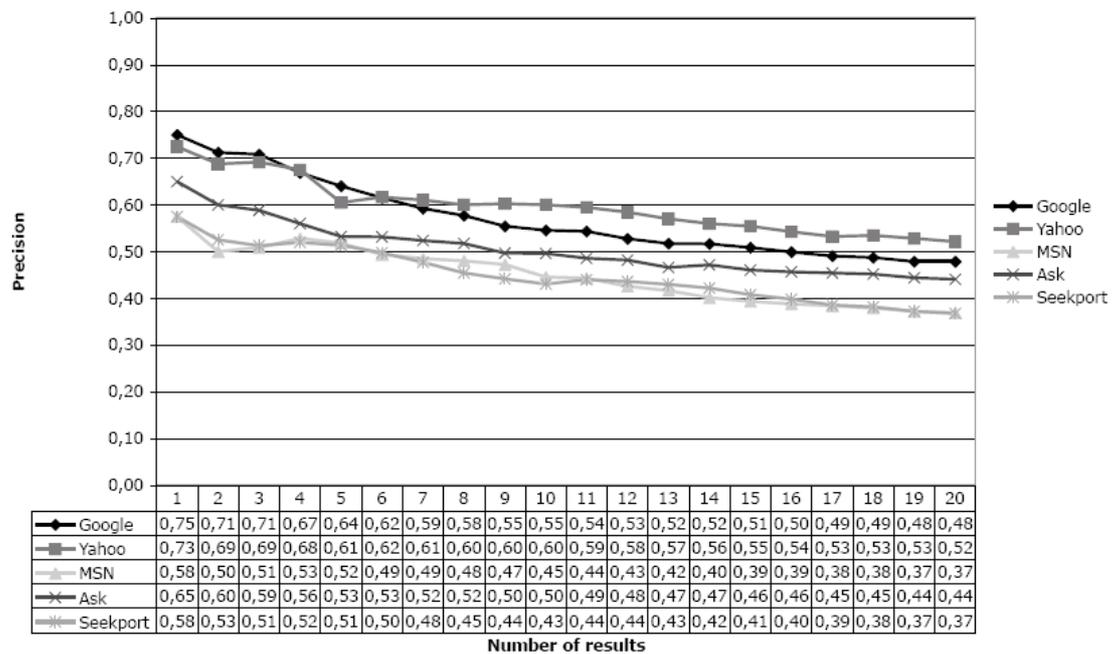

**Figure 3** Recall-precision graph (top 20 results). Ratio of relevant results in proportion to all results retrieved at the corresponding position.

*Macro precision*

Macro precision focuses on the performance of the search engines for single queries (i.e., all queries are of equal importance and the ranking position of each result is not taken into account) (see Griesbaum, 2004). We ranked the search engines according to how often they are ranked first, second, and so on for our queries [3]. Results are plotted in Figure 4.

When looking at the complete result sets (20 results per engine), we see that no single engine is able to answer all queries the best. Both the big search engines, Google and Yahoo, come first, second, or third most often, while they seldom produce bad results (and are, therefore, ranked fourth or fifth). Yahoo is able to answer 16 queries best and is, in these cases, the first choice for searchers. However, we cannot give a clear recommendation for one single search engine; the performance of the engines is largely dependent on the query submitted.

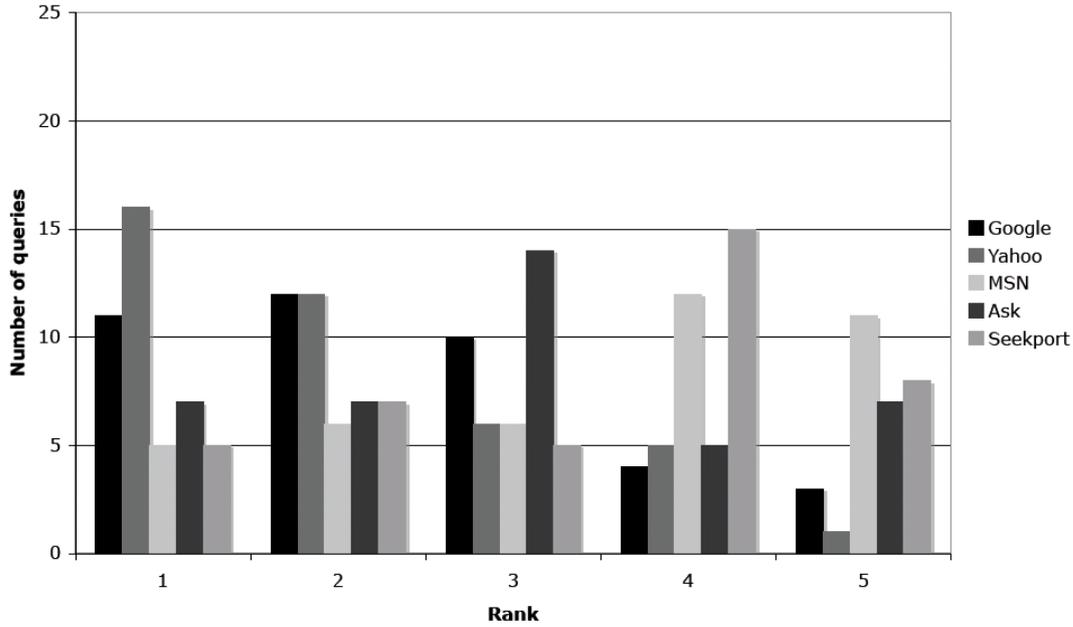

**Figure 4** Macro precision (top 20 results): Number of queries answered best, second best, … per search engine

As these results take into account all 20 results investigated, one could ask if this model reflects the real user behaviour. Therefore, we assume that a user investigates only the first three results, while the ranking of these is of no importance. We get an even clearer picture here (see Figure 5). Google and Yahoo perform best by far with more than 20 queries answered best.

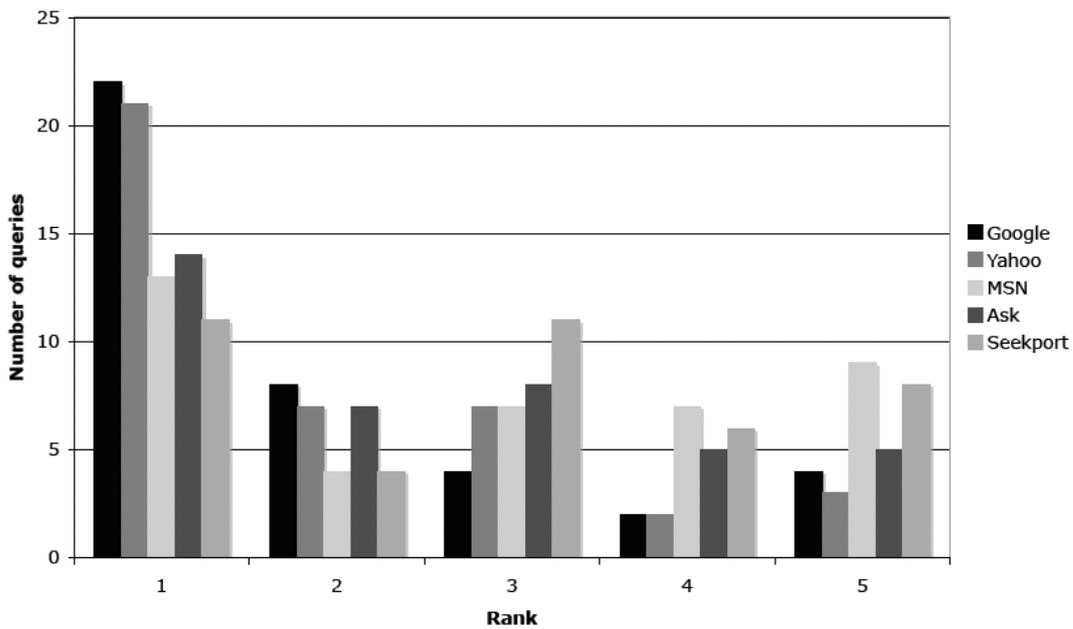

**Figure 5** Macro precision (top3 results): Number of queries answered best, second best, … per search engine

*Comparison of Results Lists vs. Results*

In this section, we compare the search engines in their ability to give adequate results descriptions (i.e., give descriptions that lead the user to relevant results). We assume that the user would only click on results with a relevant results description.

As can be seen from Figure 1, all search engines produce more relevant result descriptions than relevant results. This also holds true for individual ranks in the results ranking (i.e., the results descriptions are always judged better than the results themselves) (see Figure 6). The only exceptions are the top 1 result from MSN, Ask.com, and Seekport. Figure 7 shows the precision graphs for the results as well as for the descriptions for all engines.

To measure the distance deviance of descriptions and results, we define DRdist as the measure of description-result distance:

$$(1) \quad DRdist_n = p(description_n) - p(results_n)$$

where p is the precision measure and n is the number of results under consideration.

The results show clearly that the distance is larger for Google and MSN ($DRdist_{20}$ = 0.12 and 0.11, respectively) than for the other engines (Yahoo: 0.05; Ask.com: 0.07; Seekport: 0.07). The DRdist values do not vary greatly with the results position.

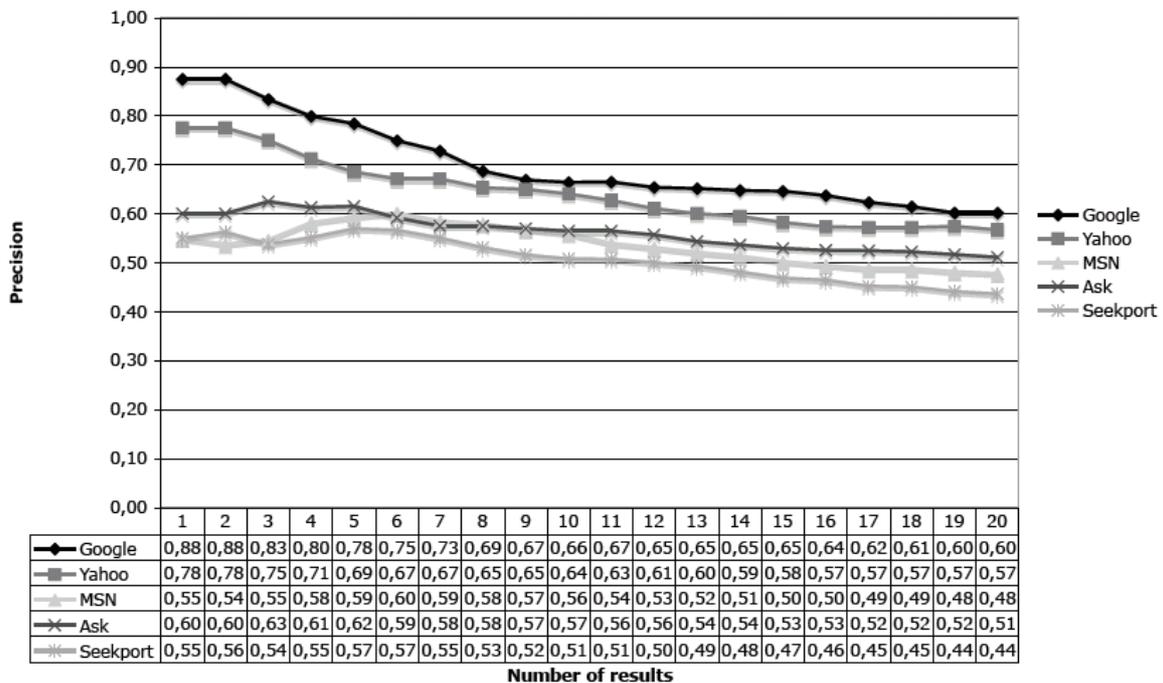

**Figure 6** Recall-precision graph (top 20 results) for results descriptions

a

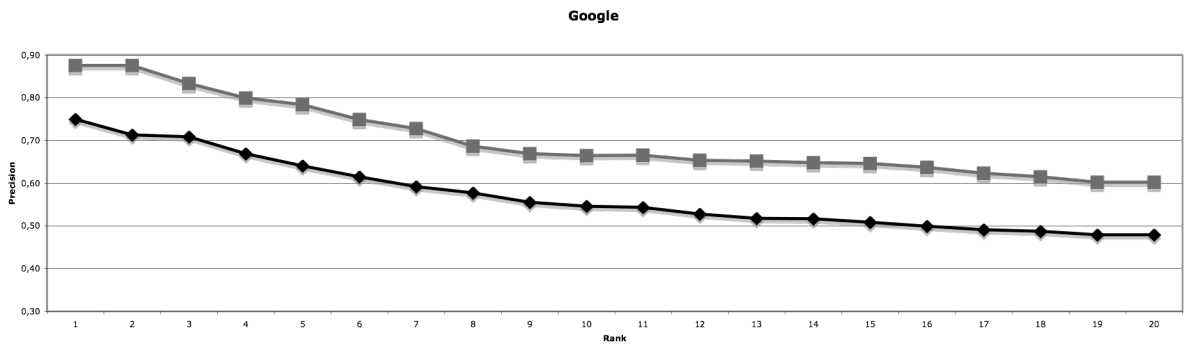

b

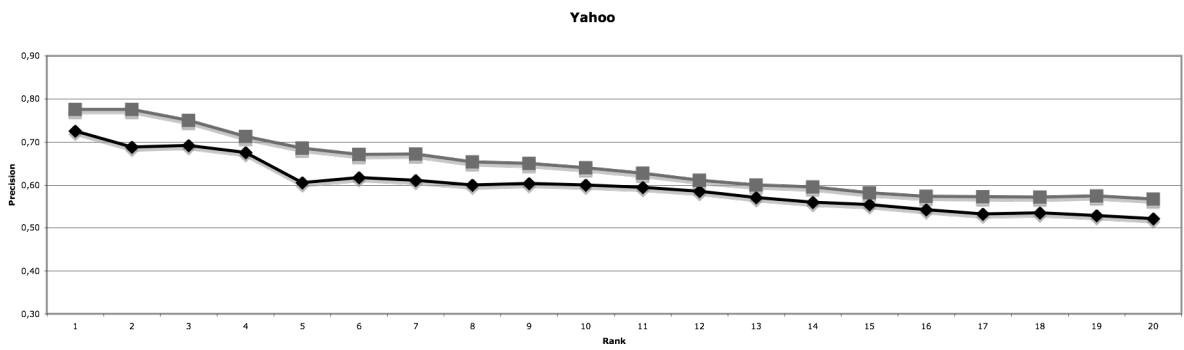

c

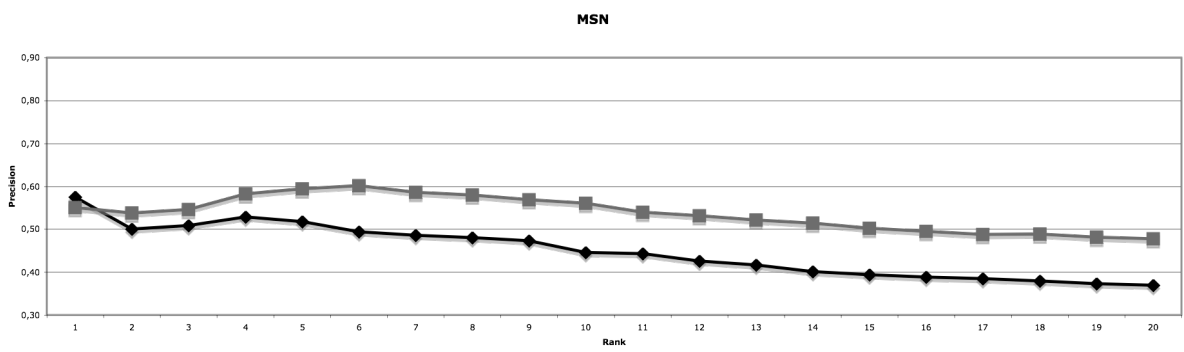

d

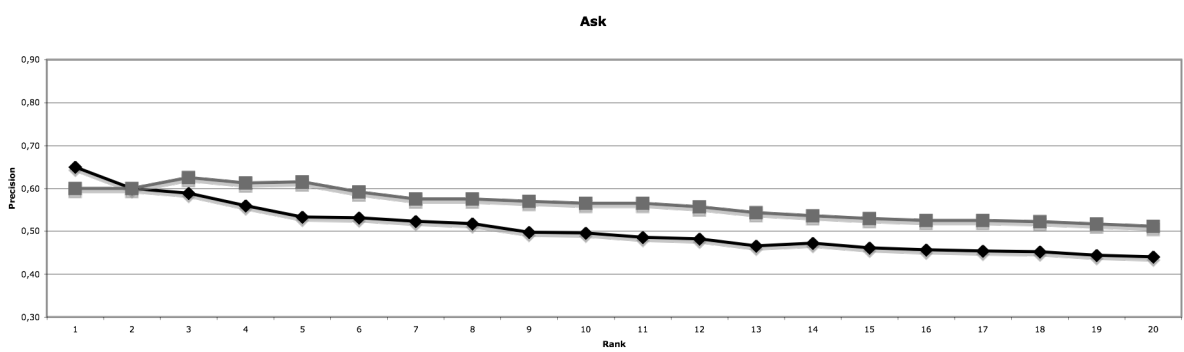

e

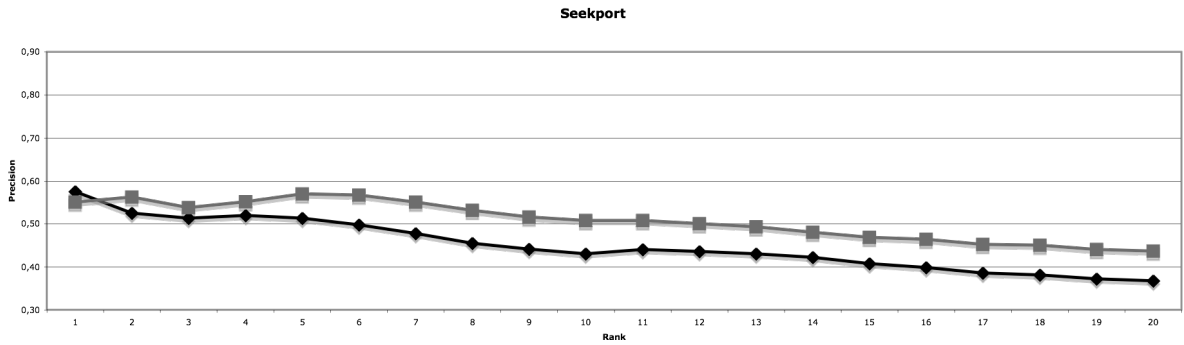

**Figure 7 (a-e)** Comparison of precision graphs (results descriptions vs. actual results) for the individual search engines

DRdist can only show to which degree the precision of descriptions and results differ, but not why this is the case and what it means for a searching person. Therefore, we will introduce further indicators that take into account the individual pairs of the precision values for descriptions and results.

However, firstly, we want to take a look at the general possibilities when comparing descriptions and results on a binary relevance scale. The four possible combinations are:

1. *Relevant description → relevant result*: This would be the ideal solution and should be provided for every result.
2. *Relevant description → irrelevant result*: In this case, the user would assume from the description that the result is relevant. The user would click on the results link. However, this would lead the user to an irrelevant result. In most cases, the user would return to the search engine results page, but frustrated to a certain degree.
3. *Irrelevant description → irrelevant result*: The user would not click on the results link because of the irrelevant description and, therefore, would not examine the irrelevant result. One could say that such results descriptions are useful in respect that at least they warn the user not to click on the irrelevant results. However, these results descriptions should not be presented on the results pages in the first place.
4. *Irrelevant description → relevant result*: In this case, the user would not consider the result because of its misleading description. The user would miss a relevant result because of its bad description.

To calculate further values, we define *e* as the total number of documents retrieved.
Table IV shows the absolute values for each indicator for the individual search engines, while Table V shows all calculations discussed in the further text.

**Table IV** Binary relevance matrix for results descriptions and actual results

|   | Description | Result | Google | Yahoo | MSN | Ask | Seekport |
|---|---|---|---|---|---|---|---|
| a | relevant | relevant | 313 | 325 | 208 | 268 | 206 |
| b | relevant | not relevant | 164 | 127 | 154 | 131 | 97 |
| c | not relevant | relevant | 67 | 90 | 73 | 76 | 51 |
| d | not relevant | not relevant | 249 | 254 | 326 | 306 | 343 |
| e | total number of documents retrieved | | 793 | 796 | 761 | 781 | 697 |

\* Different total values for individual search engines result from missing data.

**Table V** Comparison of the search engines using description-result measures
(top 20 results)

| Measure | | Google | Yahoo | MSN | Ask | Seekport |
|---|---|---|---|---|---|---|
| Description-result precision | a/e | 0,39 | 0,41 | 0,27 | 0,34 | 0,30 |
| Description-result conformance | (a+d)/e | 0,71 | 0,73 | 0,70 | 0,73 | 0,79 |
| description fallout | c/e | 0,08 | 0,11 | 0,10 | 0,10 | 0,07 |
| description deception | b/e | 0,21 | 0,16 | 0,20 | 0,17 | 0,14 |

*Description-result precision.*
The description-result precision is calculated as the ratio of results where the description as well as the result itself were judged relevant. This can be seen as a kind of "super-precision", and it should be the goal of every search engine to provide only such results. Description-results precision is calculated as follows:

$$(2) \quad DRprec = \frac{a}{e}$$

where *a* is the number of pairs of relevant descriptions and relevant results, while *e* is the total number of results.

Figure 8 shows some interesting differences compared to overall precision results (shown in Figure 3). While Google and Yahoo still perform best (differences for the top 20 results are not significant), differences between MSN and Seekport, as well as between Ask.com and Seekport, are significant. The difference between Google and Ask.com is significant only on a 0.05 level.

However, regarding only the top results, one can find clear difference between the engines. Especially on the first position, Google performs better than all other engines. This may be one reason why the general public regards Google as the best search engine. When users only observe the first few results, they may easily come to that conclusion. However, the more results are taken into account, the more the search engines adjust.

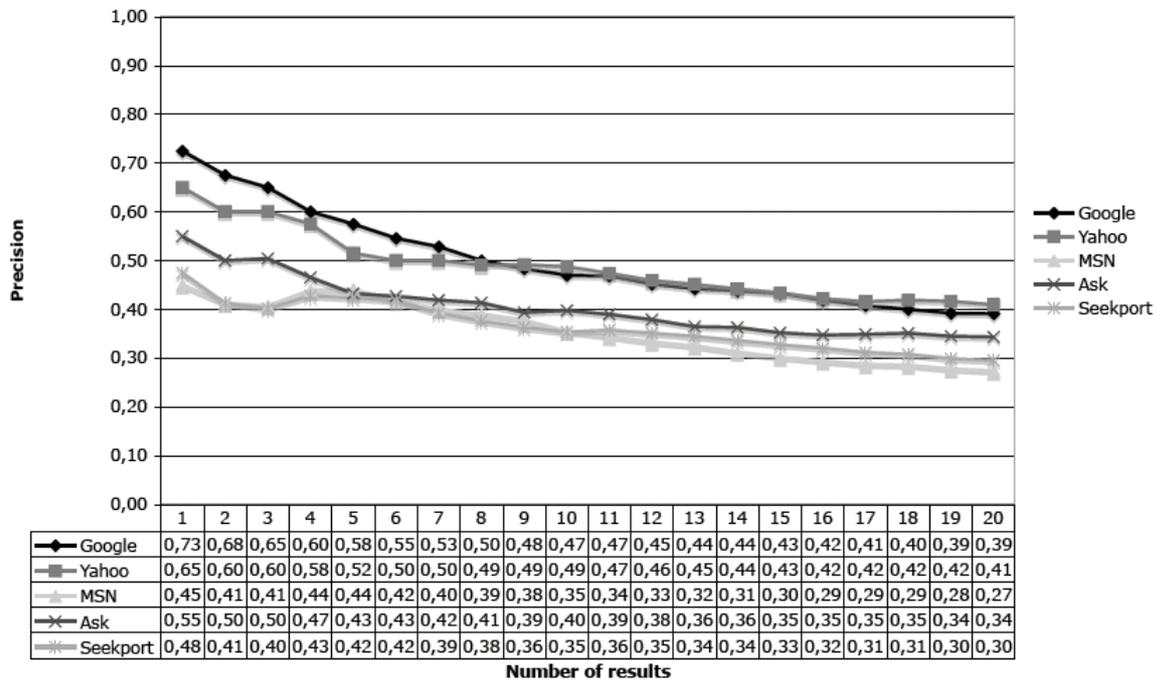

**Figure 8** Recall-DRprecision graph for top 20 results (only results where result descriptions and results are relevant)

*Description-result conformance.*
The description-result conformance takes into account all pairs where description and result are judged the same (i.e., either both are judged relevant or both are judged irrelevant). It is calculated as follows:

$$(3) \quad DRconf = \frac{a+d}{e}$$

where *a* is the number of pairs of relevant descriptions and relevant results. *d* is the number of pairs of irrelevant descriptions and irrelevant results, and *e* is the total number of results.

While description-result precision should be seen as the ideal solution, it will be helpful if the searching person can at least see from the description whether a result will be relevant of not. The disadvantage of the description-result conformance measure is that it is not only high for search engines with good DRprec values but also for the ones producing a large ratio of irrelevant results (and descriptions).

As can be seen from the DRconf values in table V, Seekport—the engine that performed worst in our test—gets the highest DRconf values (0.79). The other engines have very similar values (from 0.70 to 0.73). This shows that with the search engines under investigation, in around 70 percent of cases, a searcher can judge the results from their descriptions. However, the other way round, in around 30 percent of cases, the descriptions are misleading. Either the description appears to lead to a relevant result but in fact does not, or the description appears to lead to an irrelevant result where the result itself is relevant.

*Description fallout.*

Description fallout measures the ratio of results missed by a searcher due to a description that appears to point to an irrelevant result (i.e., the description is judged irrelevant where the result itself was judged relevant).

$$(4) \quad Dfall = \frac{c}{e}$$

where *c* is the number of pairs with irrelevant description and relevant result, while *e* is the total number of results.

In our test, description fallout values are relatively low for all search engines. Again, these values are dependent on the overall performance of the individual engines. Search engines producing completely irrelevant results and descriptions would get a high DRconf value, while Dfall (and Ddec) will be quite low.

The results show that description fallout does not appear to be a major problem for search engines. Only a relatively small number of relevant results are missed because of an unsuitable description.

*Description deception.*

Description deception measures in how many cases a user is pointed to an irrelevant result due to a description that appears to lead to a relevant result. Description deception could lead to frustrated users and should be avoided by the search engines.

Description deception is calculated as follows:

$$(5) \quad Ddec = \frac{b}{e}$$

where *b* is the number of pairs with a relevant description and a irrelevant result, while *e* is the total number of results.

As Ddec is dependent on description-result distance, the engines with the largest distance also show the highest Ddec. Google has a Ddec of 0.21 and MSN 0.20, respectively. For a user, this means that in 20 percent of cases, these engines mislead to an irrelevant result because of a description that appears relevant.

**Conclusion and Future Work**

One central question to our study was whether Google's perceived superiority in delivering relevant results could be confirmed by a systematic test. In this regard, we found that it makes no significant difference for a user to use one of the big search engines, Google or Yahoo. Only when considering solely the top three results for the pair relevant description-relevant result, can one find a slight advantage for Google. When considering the first four results, Yahoo comes close to Google, then Google leads in terms of precision for up to the seventh result. When considering more than seven results, the differences between Google and Yahoo are insignificant.

So why do users prefer Google? This may have many reasons, but we think that with this investigation, we can add one more reason previously unconsidered, namely, the results descriptions. Here, there are considerable differences between Google and Yahoo. Google delivers by far the largest amount (and the largest ratio) of relevant results descriptions. However, it is interesting to see that Google does not use techniques completely different from its competitors (all engines use KWIC in most cases), but that the identification of relevant parts of the documents used for the descriptions works better with this engine.

However, a considerable amount leads to irrelevant results. As we know from the general user behaviour with Web search engines, users do not systematically evaluate the results presented. Instead, they only click on one to a few results. We assume that the good performance of Google in user interviews (e.g., Machill et al., 2003, pp. 175-188), in part, results from the large ratio of descriptions that appear relevant.

However, the downside of the large amount of relevant results descriptions on Google is that this engine also receives a large value for description deception (i.e., users are pointed to irrelevant results more often than in other search engines [except MSN]). Search engines should work on providing the largest ratio of relevant results and, in addition, in the conformance of descriptions and results.

How should the performance of the other three search engines be judged? The relatively good performance of Ask.com may come as a surprise, as this search engine only started business in Germany in 2006. It outperforms the genuine German search engine, Seekport, as well as Microsoft's MSN, which generally disappoints.

Regarding all search engines in general, we find that their performance is far from perfect, with precision values at 20 results ranging from 0.37 to 0.52. When considering only description-result precision, the performance of the engines gets even worse (ranging from 0.27 to 0.41).

Our recommendation to the search engine companies is not only to work on improving the precision of the results but also to consider the results descriptions as well as the conformance of descriptions and results.

Our study is limited in that it is based on a model of user behaviour where the user takes into account a certain amount of results for which he evaluates at least the results descriptions. This is an advancement compared to studies that only take into account the results themselves but may not reflect the real user behaviour. Many users click on results until they find one to a few results that suit them. When they have found these, they stop their search. In future research, we want to analyse our data according to this behaviour. Additionally, we want to apply some other retrieval effectiveness measures to our data (e.g., *salience* [Ding & Marchionini, 1996]), discounted cumulated gain [Järvelin & Kekäläinen, 2002], *ability to retrieve top ranked pages* [Vaughan, 2004], and *average distance measure* [Della Mea & Mizzaro, 2004]). For an overview of promising measures for search engine evaluation, see Lewandowski (2007a). This may lead to even more accurate results.

And as a final point, we want to conduct our investigations on retrieval effectiveness more realistically in that we want to adopt our test settings to the changing results presentation in search engines (Lewandowski, 2007b). The current study should only be seen as a first step in finding appropriate measurements for retrieval effectiveness, keeping the user behaviour in mind.

**Notes**

[1] Microsoft's search engine changed its name to Live.com. However, we refer to the established name MSN, because it is the name known by the users (and still the name of the search portal).
[2] Only organic results were considered in this study. We omitted sponsored listings and "special results" (e.g., references to news results, weather forecasts) listed above the results list.
[3] When two search engines performed equal for a query, they were appointed the same rank. The next rank then was omitted, and the following search engine got the next rank. Example: SE1 gets a precision value of 0.8, SE2 0.4, SE3 0.8. Then, SE1 and SE3 get rank 1, while SE2 gets rank 3.

# References

Borlund, P. (2003). The concept of relevance in IR. Journal of the American Society for Information Science and Technology, Vol. 54, No. 10, 913-925.

Broder, A. (2002). A taxonomy of web search. SIGIR Forum, Vol. 36, No. 2, 3-10.

Buckley, C. & Voorhees, E.M. (2000). Evaluating evaluation measure stability. In Proceedings of the 23rd annual international ACM SIGIR conference on Research and development in information retrieval, Athens, Greece, ACM Press, New York, 33-40.

Burns, E. (2007). U.S. Search Engine Rankings, April 2007. Search Engine Watch. http://searchenginewatch.com/showPage.html?page=3626021

Chu, H. & Rosenthal, M. (1996). Search Engines for the World Wide Web: A Comparative Study and Evaluation Methodology. Proceedings of the 59th ASIS Annual Meeting. http://www.asis.org/annual-96/ElectronicProceedings/chu.html

Crystal, A. & Greenberg, J. (2006). Relevance Criteria Identified by Health Information Users During Web Searches. Journal of the American Society for Information Science and Technology, Vol. 57, No. 10, 1368-1382.

Della Mea, V. & Mizzaro, S. (2004). Measuring Retrieval Effectiveness: A New Proposal and a First Experimental Validation. Journal of the American Society for Information Science and Technology, Vol. 55, No. 6, 530-543.

Ding, W. & Marchionini, G. (1996). A comparative study of web search service performance. Proceedings of the 59th ASIS Annual Meeting, 136-142.

Dresel, R., Hörnig, D., Kaluza, H., Peter, A., Roßmann, N. & Sieber, W. (2001). Evaluation deutscher Web-Suchwerkzeuge. Nachrichten für Dokumentation, Vol. 52, No. 7, 381-392.

Gordon, M. & Pathak, P. (1999). Finding information on the World Wide Web: the retrieval effectiveness of search engines. Information Processing & Management, Vol. 35, No. 2, 141-180.

Granka, L.A., Joachims, T., & Gay, G. (2004). Eye-tracking analysis of user behavior in WWW search. Proceedings of the 27th Annual International ACM SIGIR Conference on Research and Development in Information Retrieval, 478-479.

Griesbaum, J. (2004). Evaluation of three German search engines: Altavista.de, Google.de and Lycos.de. Information Research, Vol. 9, No. 4. http://informationr.net/ir/9-4/paper189.html

Griesbaum, J., Rittberger, M. & Bekavac, B. (2002). Deutsche Suchmaschinen im Vergleich: AltaVista.de, Fireball.de und Lycos.de. Procedings of the 8. Internationales Symposium für Informationswissenschaft, 201-223.

Guggenheim, E. & Bar-Ilan, J. (2005). Tauglichkeit von Suchmaschinen für deutschsprachige Anfragen. Information Wissenschaft und Praxis, Vol. 56, No. 1, 35-40.

Gulli, A. & Signorini, A. (2005). The indexable Web is more than 11.5 billion pages. Proceedings of the 14th International Conference on World Wide Web, Chiba, Japan, 902-903.

Hawking, D. & Craswell, N. (2005). The Very Large Collection and Web Tracks. In TREC: Experiment and Evaluation in Information Retrieval (S. 199-231). Cambridge, Mass.: MIT Press.

Hawking, D., Craswell, N., Bailey, P. & Griffiths, K. (2001). Measuring Search Engine Quality. Information Retrieval, Vol. 4, No. 1, 33-59.

Jacsó, P. (2005). Options for presenting search results. Online Information Review, Vol. 29, No. 3, 311-319.

Järvelin, K. & Kekäiläinen, J. (2002). Cumulated gain-based evaluation of IR techniques. ACM Transactions on Information Systems, Vol. 20, No. 4, 422-446.

Lazarinis, F. (2007). Web retrieval systems and the Greek language: do they have an understanding? Journal of Information Science, in press.

Leighton, H.V. & Srivastava, J. (1999). First 20 Precision among World Wide Web Search Services (Search Engines). Journal of the American Society for Information Science, Vol. 50, No. 10, 870-881.

Lewandowski, D. (2006). Query types and search topics of German Web search engine users. Information Services & Use, Vol. 26, No. 4, 261-269.

Lewandowski, D. (2007a). Mit welchen Kennzahlen lässt sich die Qualität von Suchmaschinen messen? In Die Macht der Suchmaschinen / The Power of Search Engines (S. 243-258). Köln: von Halem.

Lewandowski, D. (2007b). Trefferpräsentation in Web-Suchmaschinen. Proceedings of Information in Wissenschaft, Bildung und Wirtschaft; 29. Online-Tagung der DGI 2007, DGI, Frankfurt am Main, in press.

Lewandowski, D. & Höchstötter, N. (2007). Web Searching: A Quality Measurement Perspective. In Web Searching: Interdisciplinary Perspectives (in press). Dordrecht: Springer.

Lewandowski, D., Wahlig, H. & Meyer-Bautor, G. (2006). The Freshness of Web search engine databases. Journal of Information Science, Vol. 32, No. 2, 133-150.

Machill, M., Neuberger, C., Schweiger, W. & Wirth, W. (2003). Wegweiser im Netz: Qualität und Nutzung von Suchmaschinen. In Wegweiser im Netz (S. 17-490). Gütersloh: Bertelsmann Stiftung.

Mizzaro, S. (1997). Relevance: The Whole History. Journal of the American Society for Information Science, Vol. 48, No. 9, 810-832.

Rose, D.E. & Levinson, D. (2004). Understanding user goals in Web search. Proceedings of the thirteenth International World Wide Web Conference, ACM Press, New York, 13-19.

Schmidt-Maenz, N. & Koch, M. (2006). A General Classification of (Search) Queries and Terms. Proceedings of the 3rd International Conference on Information Technologies: Next Generations, Las Vegas, Nevada, USA, 375-381.

Spink, A., Wolfram, D., Jansen, M. & Saracevic, T. (2001). Searching the Web: The public and their queries. Journal of the American Society for Information Science and Technology, Vol. 53, No. 2, 226-234.

Stock, M. & Stock, W.G. (2000). Internet-Suchwerkzeuge im Vergleich, Teil 1: Retrievaltest mit Known Item Searches. Password, Vol. 15, No. 11, 23-31.

Sullivan, D. (2006). Hitwise Search Engine Ratings. Search Engine Watch. http://searchenginewatch.com/showPage.html?page=3099931

Sullivan, D. (2007). Major Search Engines and Directories. Search Engine Watch. http://searchenginewatch.com/showPage.html?page=2156221

Tague-Sucliffe, J. (1992). The pragmatics of information retrieval experimentation, revisited. Information Processing & Management, Vol. 28, No. 4, 467-490.

Vaughan, L. (2004). New Measurements for Search Engine Evaluation Proposed and Tested. Information Processing & Management, Vol. 40, No. 4, 677-691.

Vaughan, L. & Thelwall, M. (2004). Search Engine Coverage Bias: Evidence and Possible Causes. Information Processing & Management, Vol. 40, No. 4, 693-707.

Véronis, J. (2006). A comparative study of six search engines. http://www.up.univ-mrs.fr/veronis/pdf/2006-comparative-study.pdf

Webhits Web-Barometer (2006). http://www.webhits.de/deutsch/index.shtml?webstats.html

Wolff, C. (2000). Vergleichende Evaluierung von Such- und Metasuchmaschinen. Proceedings of 7. Internationales Symposium für Informationswissenschaft, Darmstadt, Germany, Universitätsverlag, Konstanz, 31-48.